# Acoustic metamaterials with spinning components

Degang Zhao[1,2,6], Yao-Ting Wang[2,4,5,6], Kin-Hung Fung[3], Zhao-Qing Zhang[2], and C. T. Chan[2]★

**Using both multiple scattering theory and effective medium theory, we find that an acoustic metamaterial consisting of an array of spinning cylinders can possess a host of unusual properties including folded bulk and interface-state bands in the subwavelength regime. The folding of the bands has its origin in the rotation-induced antiresonance of the effective compressibility with its frequency at the angular velocity of the spinning cylinders, as well as in the rotational Doppler effect which breaks the chiral symmetry of the effective mass densities. Both bulk and interface-state bands exhibit remarkable variations as the filling fraction $f$ of the spinning cylinders is increased. In particular, a zero-frequency gap appears when $f$ exceeds a critical value. The uni-directional interface states bear interesting unconventional characteristics and their robust one-way transport properties are demonstrated numerically.**

Acoustic metamaterials (AMs) are artificial acoustic composites made of subwavelength microstructures and they are designed to exhibit novel wave phenomena or functionalities not found in natural acoustic materials. Exotic examples of these phenomena include acoustic superlenses [1-3], invisibility cloaks [4-8], acoustic black holes [9-13] and topological acoustic materials [25-29]. One of the most striking properties of AMs is that the effective constitutive parameters can become negative in the low-frequency regime. For example, negative mass density can be achieved by a composite consisting of heavy metal beads coated in soft rubber [14,15], while negative bulk modulus can be achieved by a periodic array of Helmholtz resonators [16]. Subsequently, various AMs simultaneously possessing negative mass density and negative bulk modulus have been proposed [17-20]. Most AMs exhibiting negative parameters derive their functionality from the oscillatory motions of the inclusions driven by the impinging wave. If one allows relative motion between the constituent materials in AMs, the additional degrees of freedom due to kinetic equilibrium can give rise to even more interesting phenomena. As noted by Censor *et al.* [21], sound waves can be scattered by an isotropic scatterer rotating at a constant angular velocity even if the spinning inclusion and the background medium are identical. Their results imply that the rotation of the scatterer can dramatically transform the acoustic properties of an AM. Furthermore, the assumption of rotation of inclusions implicitly implies the existence of interactions between the system and the environment driving the inclusions at a pre-specified angular velocity. The system is in fact non-Hermitian. Recently, the non-Hermitian acoustic system has attracted much attention thanks to its many unique properties that are absent in conventional Hermitian systems [22-29].

[1]School of Physics, Huazhong University of Science and Technology, Wuhan 430074, China. [2]Department of Physics, The Hong Kong University of Science and Technology, Clear Water Bay, Hong Kong, China. [3]Department of Applied Physics, The Hong Kong Polytechnic University, Hong Kong, China. [4]Department of Physics, Imperial College London, London, SW7 2AZ, United Kingdom. [5]School of Physics and Astronomy, The University of Birmingham, Birmingham, B15 2TT, United Kingdom. [6]These authors contributed equally: Degang Zhao, Yao-Ting Wang. ★e-mail:phchan@ust.hk

In this work, we study both the bulk bands and the interface-state bands of AMs consisting of spinning cylinders using multiple scattering theory (MST) [30,31] as well as effective medium theory (EMT) [32]. Both the bulk and interface-state bands exhibit interesting band-folding behaviors in the subwavelength realm, corresponding to the so-called "k-gaps" typically found in non-Hermitian systems [35, 42-45]. The effective compressibility shows a rotation-induced anti-resonance behavior [41], which induces a low-frequency gap at a high enough filling ratio. The presence of this antiresonance, together with the rotational Doppler effect which breaks the chiral symmetry of the effective mass densities, leads to the formation of the folded bulk and interface-state bands. The interface states are unidirectional and their robust one-way transport is numerically demonstrated. The evolution of bulk and interface-state bands is systematically analyzed by varying the filling fraction $f$ of the spinning cylinders in the lattice. We find that the band dispersion can change dramatically when the filling fraction is varied. This rotation-induced dispersion enables us to realize a new class of AMs with unique properties.

We begin by considering a two-dimensional (2D) periodic array of solid circular cylinders, each with mass density $\rho_s$ and acoustic wave velocity $c_s$, immersed in a fluidic background with mass density $\rho_0$ and acoustic wave velocity $c_0$, as shown in Fig. 1(a). We assume that the impedance contrast between the cylinders and the background medium is large so that shear waves in the cylinders can be ignored [39,40]. The lattice constant is $a$, the radius of cylinders is $r_s$ and each cylinder rotates about its symmetric $z$ axis at a uniform angular velocity $\Omega$, which is defined as positive (negative) for anticlockwise (clockwise) rotation. While Fig. 1(a) illustrates a square lattice of cylinders, the lattice symmetry is in fact immaterial because we are interested in the low-frequency regime where the material can be viewed as a homogenized medium. For simplicity, we ignore the viscosity of the background medium in our theoretical analysis. In polar coordinates, the radial component of the pressure acoustic wave equation inside the inclusion is given by [21]

$$r^2\partial_r^2 p + r\partial_r p + (r^2\lambda_m^2 - m^2)p = 0, \tag{1}$$

where $p$ is the pressure field; $m$ is the integer angular index, which yields $\partial_\theta p = imp$; and $\lambda_m^2 = -(4\Omega^2 + M^2)c_s^{-2}$ gives the rotating wave number $\lambda_m$. The frequency correction $M = -i(\omega - m\Omega)$ arises from the rotational Doppler effect [33] that causes frequency shifts of waves propagating from static to rotating media and breaks the symmetry between $m$ and $-m$ states. Outside the inclusion, Eq. (1) is replaced by

$$r^2\partial_r^2 p + r\partial_r p + (r^2 k_0^2 - m^2)p = 0, \tag{2}$$

where $k_0 = \omega/c_0$ denotes the wave number of the background medium. The general solutions of Eqs. (1) and (2) have the forms $p_{in} = \sum_{m=-\infty}^{\infty} a_m J_m(\lambda_m r)e^{im\theta}$ inside the inclusion and $p_{out} = \sum_{m=-\infty}^{\infty} b_m J_m(k_0 r)e^{im\theta} + c_m H_m(k_0 r)e^{im\theta}$ outside the inclusion. Here $J_m(x)$ and $H_m(x)$ are the Bessel function and Hankel function of the first kind, respectively. To match the boundary conditions, we further need the radial displacements $u_r$ inside the inclusion, which has the form [21]

$$u_r = \left[(2\Omega^2 - M^2)\partial_r p - 3imM\Omega p/r\right] / \left[\rho_s(M^2 + 4\Omega^2)(M^2 + \Omega^2)\right]. \tag{3}$$

After applying the boundary conditions that both $p$ and $u_r$ are continuous at $r = r_s$, the Mie scattering coefficients can be obtained as [21]

$$D_m = -\frac{\lambda_m \rho_0 R_m(\lambda_m r_s)J_m(k_0 r_s) - k_0 \rho_s J_m(\lambda_m r_s)J'_m(k_0 r_s)}{\lambda_m \rho_0 R_m(\lambda_m r_s)H_m(k_0 r_s) - k_0 \rho_s J_m(\lambda_m r_s)H'_m(k_0 r_s)}, \tag{4}$$

where

$$R_m(\lambda_m r_s) = \frac{\omega^2}{(4\Omega^2 + M^2)(\Omega^2 + M^2)}\left[(2\Omega^2 - M^2)J'_m(\lambda_m r_s) - 3imM\Omega\frac{J_m(\lambda_m r_s)}{\lambda_m r_s}\right]. \tag{5}$$

Eq. (4) describes the scattering of a single rotating cylinder. This equation enables us to calculate the band structures of the system by using MST [30, 31] in any frequency range as well as the effective medium parameters in the long-wavelength limit.

Knowing the Mie scattering coefficients $D_m$, we can obtain the effective medium parameters of the homogenized medium in the long-wavelength regime. Using the EMT technique [32], we obtain the following effective bulk modulus $B_e$ and effective mass densities $\rho_e^\pm$ from $D_0$ and $D_{\pm 1}$, respectively (see Section I of Supplementary Information for details):

$$B_e^{-1} = fB_{es}^{-1} + (1-f)B_0^{-1} \quad \text{with} \quad B_{es}^{-1} = \frac{2\Omega^2 + \omega^2}{\rho_s c_s^2 (\omega^2 - \Omega^2)}$$

$$\rho_e^\pm = \rho_0 \frac{(1+f)\rho_{es}^\pm + (1-f)\rho_0}{(1-f)\rho_{es}^\pm + (1+f)\rho_0} \quad \text{with} \quad \rho_{es}^\pm = \rho_s \left(1 \pm \frac{\Omega}{\omega}\right) \tag{6}$$

where $f$ denotes the filling fraction of the cylinders and the $\pm$ sign refers to anticlockwise/clockwise rotation ("$m = \pm 1$"). $B_{es}$ and $\rho_{es}^\pm$ are, respectively, the effective bulk modulus and effective mass densities of the spinning cylinders seen in the laboratory frame. The $B_{es}$ and $\rho_{es}^\pm$ expressions can also be independently derived from the scattering of a single spinning cylinder. (See Eqs. (S28), (S29) and (S40) in Supplementary Information). The splitting of $\rho_e^+$ and $\rho_e^-$ (or $\rho_{es}^+$ and $\rho_{es}^-$) is a result of the rotational Dopple effect, which makes $D_1 \neq D_{-1}$. Eq. (6) indicates that an ordinary material that has a frequency-independent response can be turned into a dispersive medium when parts of the medium are spinning. In addition, transforming the density terms of Eq. (6) into Cartesian coordinates in the x-y plane leads to an anisotropic density tensor of the form (see Section II of Supplementary Information for details)

$$\overleftrightarrow{\rho}_e = \frac{1}{2}\begin{pmatrix} \rho_e^+ + \rho_e^- & -i(\rho_e^+ - \rho_e^-) \\ i(\rho_e^+ - \rho_e^-) & \rho_e^+ + \rho_e^- \end{pmatrix}. \tag{7}$$

By substituting $B_e$ and $\overleftrightarrow{\rho}_e$ into the acoustic wave equation

$$B_e \nabla(\nabla \cdot \vec{v}) = \overleftrightarrow{\rho}_e \frac{\partial^2 \vec{v}}{\partial t^2}, \tag{8}$$

we obtain the following effective medium dispersion relation:

$$k^2 = \frac{2\omega^2 \rho_e^+ \rho_e^-}{(\rho_e^+ + \rho_e^-)B_e}. \tag{9}$$

As a concrete example, we consider a prototypical system in which the rotating cylinders are epoxy ($\rho_s = 1180\,\text{kg/m}^3, c_s = 2540\,\text{m/s}$) and the background medium is water

($\rho_0 = 1000\,\text{kg/m}^3$, $c_0 = 1490\,\text{m/s}$) and we set $a = 0.1\,\text{m}$, $f = 0.3$ and $\Omega = 3000\,\text{rad/s}$. The band structures calculated using MST and EMT are compared in Fig. 1(b). We emphasize here that similar results can be obtained for other material parameters. The excellent agreement between the EMT and MST results confirms the validity of Eq. (9). It is interesting to find that two rotation-induced band gaps emerge in the low-frequency regime with a folded band lying in between. The physics of these resonant gaps and the folded band can be understood from the dispersion of the effective parameters depicted in Fig. 1(c). According to Eq. (9), a band gap occurs if its right-hand side is negative, which is determined by the signs of $B_e^{-1}$, $\rho_e^+$, $\rho_e^-$ and $\rho_e^+ + \rho_e^-$. Fig. 1(c) clearly indicates that the first band gap stems from the single negativity of $\rho_e^-$. The second band gap originates from the negativity of the effective compressibility $B_e^{-1}$ arising from the antiresonance of $B_e^{-1}$ at $\omega = \Omega$ as the frequency dependence of the response function $B_e^{-1}$ exhibits an opposite trend to that of $\rho_e^-$ [41]. Thus, the presence of antiresonance in $B_e^{-1}$, together with the rotational Doppler effect which breaks the symmetry between $\rho_e^+$ and $\rho_e^-$, produces the folded band between the two gaps. The folded band undergoes a transition from positive to negative group velocity as the frequency is increased. At the transition frequency, the group velocity diverges and the average group refractive index vanishes, i.e., $n_g = \partial(n\omega)/\partial\omega = 0$ [35]. (See Section III of Supplementary Information). This point can be regarded as an exceptional point associated with the coalescence of two modes and the opening of a "k-gap" [42, 45-48]. It should be noted that the folded bands in photonic systems [35, 42, 48] obtained previously are due to the assumption of non-dispersive negative constitutive parameters, which is incompatible with causality, while the rotation-induced folded band here originates from a realistic model.

Since the zeros of the functions $\rho_e^-(\omega)$ and $B_e^{-1}(\omega)$ depend on the filling fraction $f$, the frequency range of the folded band as well as the entire band structure can vary drastically with a change in the filling fraction. To investigate the variations in band structure induced by a changing filling fraction, it is convenient to consider the frequencies of the four band-edge states at the Brillouin zone boundaries (marked by $\omega_1$, $\omega_2$, $\omega_3$ and $\omega_4$ in Fig. 1(b)), which can be obtained explicitly from Eq. (9). At $k = 0$, $\omega_1$ and $\omega_2$ are determined by the conditions of

$\rho_e^-(\omega_1) = 0$ and $B_e^{-1}(\omega_2) = 0$. For $\omega_3$ and $\omega_4$, for convenience, we take $k = \infty$ in Eq. (9) and look for the conditions of $\rho_e^+(\omega_3) + \rho_e^-(\omega_3) = 0$ and $B_e^{-1}(\omega_4) = \infty$, respectively. The analytical solutions of $\omega_1$, $\omega_2$, $\omega_3$ and $\omega_4$ can be found in Supplementary Information (Section IV) and their dependence on $f$ is shown in Fig. 2(a). We note that $\omega_4 = \Omega$ is independent of $f$. It is interesting to see that $\omega_2$, as a rapidly decreasing function of $f$, intersects $\omega_1$ at $f = 0.3887$, at which the folded band shrinks to a point (Fig. 2(b)). The folded band re-emerges as $f$ is further increased, but is inverted with $\omega_2$ lying below $\omega_1$ (Fig. 2(c)). When $f$ approaches $0.5516$, $\omega_2 \cong \omega_3$ with the two lowest bands almost touching (Fig. 2(d)). When $f$ is above $0.5516$, $\omega_2$ lies below $\omega_3$ and the first band becomes folded (Fig. 2(e)). Finally, $\omega_2$ approaches zero when $f \approx 0.6316$, above which the first band disappears and the system becomes gapped with a cutoff frequency at $\omega_3$ (Fig. 2(f)). The effective parameters at different $f$ are shown in Fig. S2 of Supplementary Information. The presence of a zero-frequency gap has its origin in the negativity of $B_e^{-1}$ at low frequencies when $f$ is sufficiently large as shown in Fig. S2(e). In fact, it can be shown that a spinning cylinder has a negative response to an external applied force at a low frequency due to the anti-resonant nature of $B_{es}^{-1}$ shown in Eq. (6) (detailed analysis can be found in Section V of Supplementary Information).

In a 2D system, the acoustic wave equation has the same form as the electromagnetic wave equation (for one specific polarization) in the absence of rotation, and the acoustic and electromagnetic parameters have a one-to-one mapping, $1/B \leftrightarrow \varepsilon, \rho \leftrightarrow \mu$ when the electric field is parallel to the cylinder axis. In the presence of rotation, according to Eq. (7) our model is similar to the gyromagnetic photonic crystal under an external dc magnetic field, which induces the imaginary off-diagonal terms in $\mu$ [36, 37]. As such, the spinning of cylinders plays the counterpart of an effective static magnetic field, which breaks the time-reversal symmetry and reciprocity, mimicking a quantum Hall system under an external magnetic field. If we combine the lattice with anticlockwise spinning cylinders and that with clockwise spinning cylinders (mimicking the magnetic field applied antiparallel to the gyromagnetic photonic crystal), we would expect the interface states to exist in the subwavelength band gap. However, there is one important difference. While the presence of the magnetic field changes $\mu$ without affecting the electric permittivity $\varepsilon$, the rotation of cylinders in our system not only produces an antiresonance in the effective compressibility, it also makes the system non-Hermitian, leading to the folding of the bulk bands shown in Figs. 1(b) and 2(b-f). As we will show below, this antiresonance can also fold an interface-state band so that the interface states can propagate in two opposite directions in two different portions of an interface band.

We now investigate the dispersion relation of interface states. We consider two semi-infinite media separated by a boundary at $y=0$, each having spinning cylinders at the same filling fraction but whose angular velocities display different signs ($\Omega$ for $y<0$ and $-\Omega$ for $y>0$). The pressure of the interface states takes the form $p = p_0 e^{ik_x x - \beta|y| - i\omega t}$ and by applying EMT, the propagation constants can be solved as

$$k_x^2 = \frac{\left(\rho_e^+ + \rho_e^-\right)\omega^2}{2B_e} \quad \text{with} \quad \beta = -\frac{\rho_e^+ - \rho_e^-}{\rho_e^+ + \rho_e^-}k_x > 0. \tag{10}$$

Since $\beta$ is positive in Eq. (10), these interface states are unidirectional. The sign of $k_x$ is determined by the relative magnitudes of $\rho_e^+$ and $\rho_e^-$ and can therefore have different signs in different bands. Moreover, an additional non-dispersive flat band occurs at frequency $\omega_1$ where $\rho_e^- = 0$, i.e.,

$$\rho_e^- = 0 \quad \text{with} \quad \beta = k_x > 0. \tag{11}$$

This flat band has zero group velocity and does not transport energy. Similar flat bands were found previously at the interface of two magnetic domains of some Hermitian system [38]. The detailed derivations of the dispersion relations of interface states and the flat band can be found in Section VI of Supplementary Information.

We now use MST to numerically verify the interface state dispersion derived above. We construct a supercell with 10 cylinders rotating anticlockwise in the region $y<0$ and 10 cylinders rotating clockwise in the region $y>0$. In the MST calculation, we apply the periodic boundary condition on the two edges of the supercell. For the case of $f = 0.3$, the result of the four lowest interface-state bands calculated by MST are shown by the blue circles in Fig. 3(a). Also plotted in Fig. 3(a) is the result of EMT given in Eqs. (10) and (11) (red solid lines) with the effective parameters shown in Fig. 3(b). Again, excellent agreement is found. Unlike the projected bulk bands (green shaded region), where the dispersions are symmetric for the positive and negative $k$ solutions, interface bands appear only on one side of $k_x$ determined by the condition of $\beta > 0$ according to Eqs. (10) and (11). Thus, the sign of $k_x$ can vary from band to band as shown in Fig. 3(a). For the three dispersive interface bands shown in Fig. 3(a), only a portion of the second band lies within the absolute gap of the projected bulk bands. The lowest and the third interface bands overlap spectrally with the corresponding bulk bands. Similar to the case of the folded bulk band, we find also a transition point at which the group velocity diverges and the average group refractive index for the interface states vanishes, i.e., $n_g^{(s)} = 0$ (see Section III of Supplementary Information). The third band is also folded and this band lies very close to the bulk band due to the fact that $\rho_e^+ - \rho_e^- \to 0$ at high frequencies as clearly depicted in Fig. 3(b). As a result, the decay length diverges and the dispersion relations of the interface states and bulk states merge as shown in Eqs. (10) and (9).

Using the EMT result of Eq. (10), we can analytically solve the frequencies of interface states on the Brillouin zone boundaries, marked by $\omega_1'$, $\omega_2'$, $\omega_3'$ and $\omega_4'$ in Fig. 3(a) (see Section VII of Supplementary Information). All of these frequencies are also functions of the filling fraction $f$, which is again an important parameter that determines the pattern of interface-state bands, whose evolution is demonstrated in Fig. S4 in Section VII of Supplementary Information. To the best of our knowledge, the unusual dispersion of interface states and their sensitive dependence on the filling fraction are not found in Hermitian systems. To explicitly demonstrate the interface states, we show in Figs. 3(c) and (d) the amplitude of the pressure field of the two eigenstates calculated by MST: one is in the first band and the other is in the second band. Both states exhibit an exponential decay behavior away from the interface. To determine the decay coefficient $\beta$, in Figs. 3(e) and (f), we plot the magnitude of the average field inside the cylinders on one side of the supercell as a function of the distance away from the interface. The results of exponential fittings give the values of $\beta$ of these two states as 3.06 and 1.81 respectively, which agree well with the values of 3.04 and 1.79 obtained from Eq. (10).

To demonstrate the robustness of the one-way transport of the interface states, we choose the filling fraction $f = 0.3887$ as an example. The corresponding projected bulk bands and the interface states are shown in Fig. 4(a). In this case, since the second folded bulk band shrinks almost to a point, a wider gap exists in the bulk band structure, making it easier to observe the robust one-way transport of interface states. Unlike the conventional topological interface-state band which typically has either a positive or a negative group velocity [25-29] within one band, our "folded" interface band possesses both positive and negative group velocities. The unidirectional transmission of interface states along two different directions can be achieved at different frequencies in the same band. A finite-sized system that has a boundary with rectangular U-turns is purposely designed, as shown in the inset of Fig. 4(b). The cylinders above the interface rotate in the clockwise direction while those below the interface rotate in the anticlockwise direction and the entire system is embedded in a static water background. Since surface states may also exist on the surfaces between the rotating media and the static background, they are eliminated by adding absorption to the surrounding layer. A 2D point source is placed at the center of the structure (marked by the red star in Fig. 4(b)). The frequency of the transition point in the folded interface-state band that marks the change from a positive to a negative group velocity is $\omega a/2\pi c_0 = 0.0202$. Fig. 4(b) clearly demonstrates that the excited interface wave propagates along the left side of the interface at frequency $\omega a/2\pi c_0 = 0.0206$ and along the right side at frequency $\omega a/2\pi c_0 = 0.0211$, without backscattering from the sharp corners. The directions of propagation are consistent with the signs of the group velocity shown in Fig. 4(a).

Finally, we discuss possible experimental realization of the proposed model. In an actual experiment, spinning rods can be controlled by external agents such as electric motors. The rotatational speed for our examples lies in the range of 28k-38k rpm or 3k-4k rad/s and this range should be operationally feasible. Slower rotational speeds will yield similar results, except that the band gaps will become narrower. In the present study, we have considered cylinders with circular cross-sections. The system is identical across time points and hence we can use the standard band structure to describe the physics. A natural extension is to consider cylinders with a corrugated boundary or cylinders with lower symmetry (e.g. ellipse). In these cases, we will need to consider the Floquet bands as the system is different at each time point and we expect more fascinating physics to emerge. Another direction worth investigating is whether the interesting rotation-induced effects can also be observed when elastic waves are considered. These effects include the effective bulk modulus and the mass density tensor becoming dispersive; a rotation-induced low-frequency cutoff for high filling ratios; and antiresonance behavior in the effective compressibility.

This work was supported by the Hong Kong Research Grants Council under grant no. AoE/P-02/12. Degang Zhao was also supported by the National Natural Science Foundation of China under grant no. 11874168.

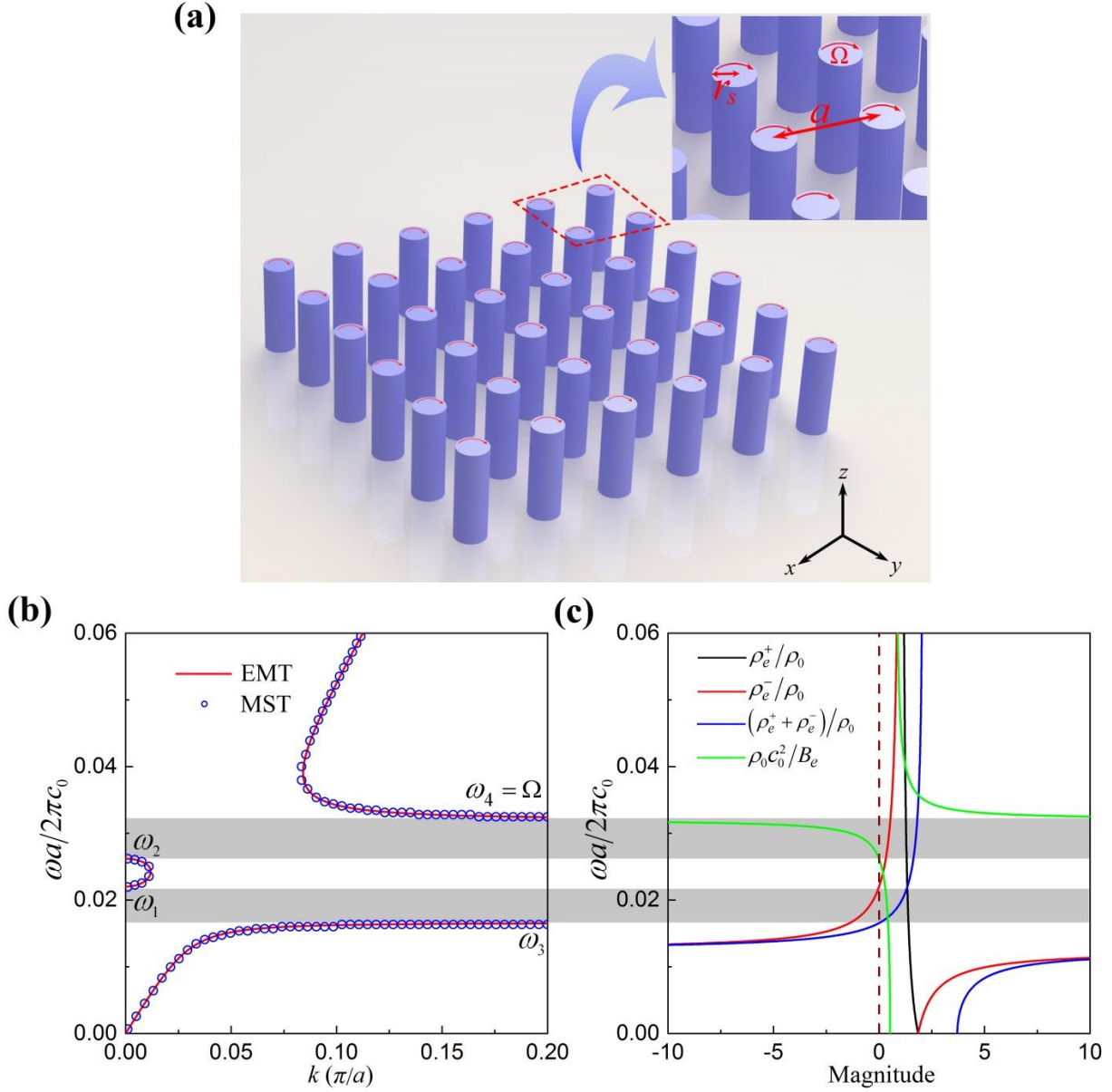

Fig. 1 (a) A two-dimensional square array of rotating solid cylinders in a fluidic background. Each cylinder rotates about its symmetric axis at a constant angular velocity Ω. (b) The band structures calculated by MST and EMT at a filling ratio $f = 0.3$ showing two rotation-induced gaps (shaded in grey) sandwiching a "folded" band in between. The bands intercept the Brillouin zone center and boundaries at $\omega_1$, $\omega_2$, $\omega_3$ and $\omega_4$. (c) Effective mass density and compressibility versus frequencies at $f = 0.3$, showing that the first and second gaps orginate respectively from negative density and negative modulus induced by rotation.

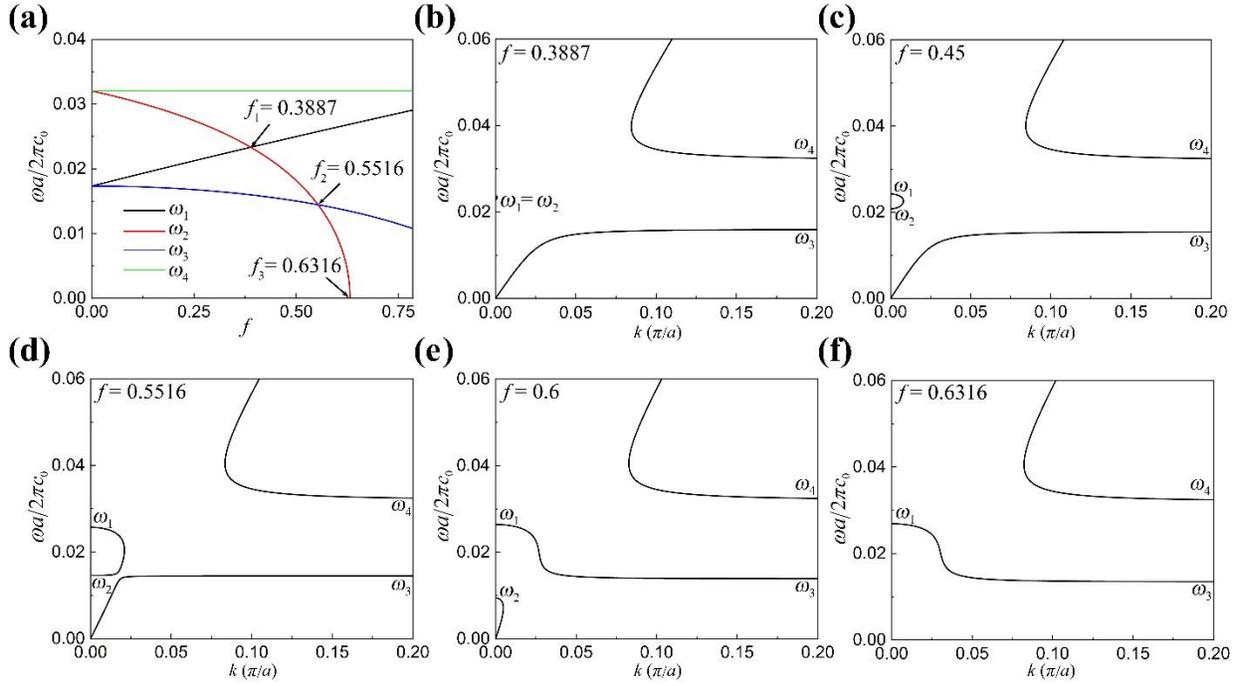

Fig. 2 (a) Frequencies of four band-edge states versus filling fraction $f$. (b)-(f) Evolution of bulk bands with increasing $f$. When $f > 0.6316$, $\omega_3$ becomes a cutoff frequency below which acoustic waves cannot propogate.

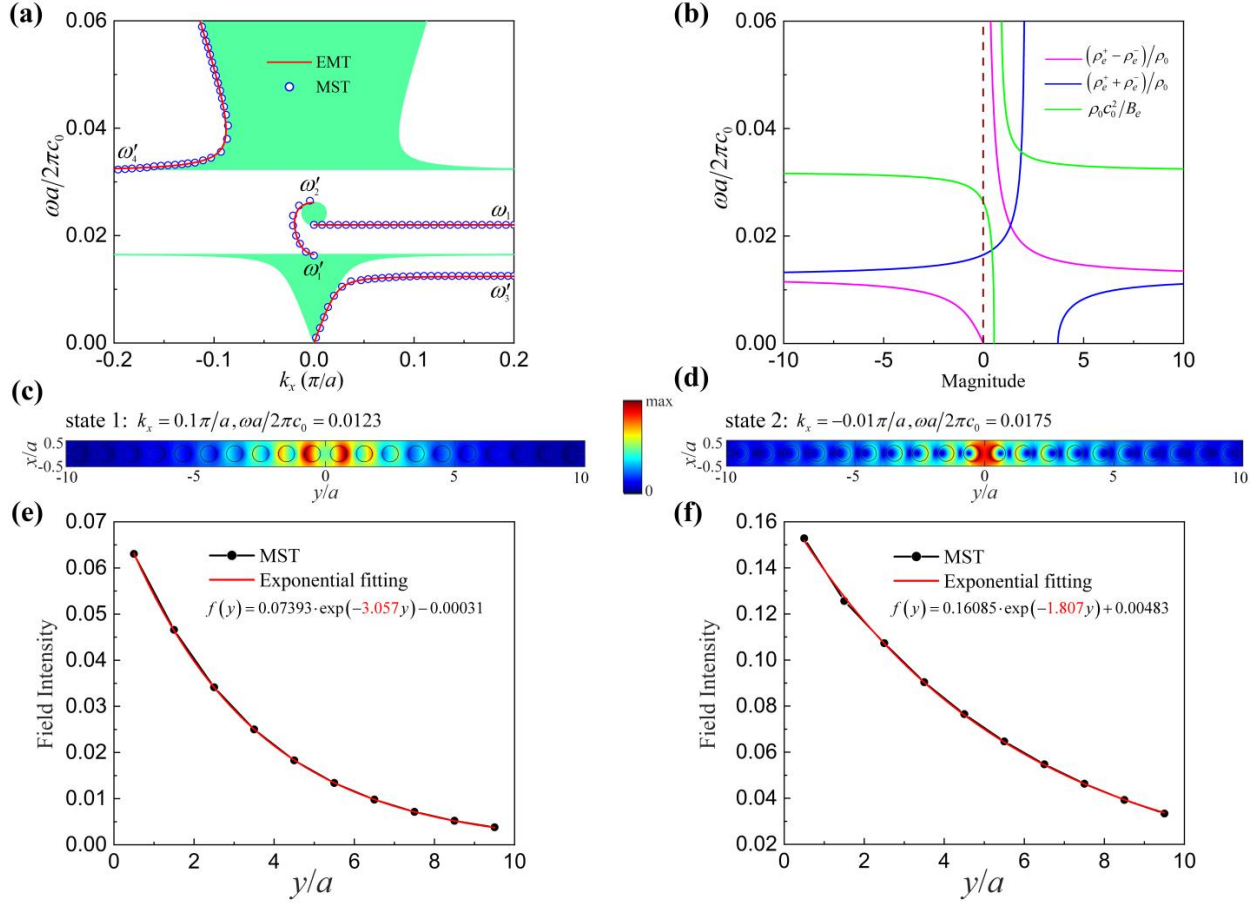

Fig. 3 (a) Interface state dispersion relations calculated by MST and EMT at $f = 0.3$. $\omega_1$ is the frequency of the flat band. $\omega'_1$, $\omega'_2$, $\omega'_3$ and $\omega'_4$ denote the frequencies at the Brillouin zone center and boundaries. The green shaded regions represent the projected bulk bands along the $x$ direction. (b) Effective mass densities and compressibility. (c) and (d) are the absolute values of the pressure field distribution of two eigenstates in the first and second interface-state bands, respectively. (e) and (f) are the intensities of the average field inside all cylinders on the $y > 0$ side for the two states in (c) and (d), respectively. The exponential fittings are also presented.

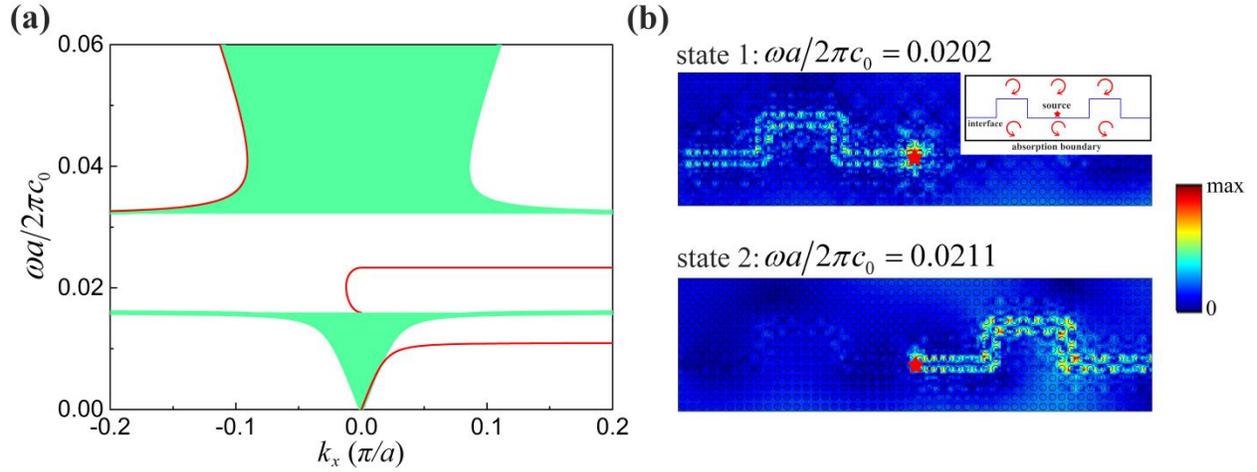

Fig. 4 (a) Dispersion relation of projected bulk band (green shaded region) and interface states (red solid line) with a filling fraction $f = 0.3887$. (b) The wave with frequency $\omega a/2\pi c_0 = 0.0202$ and that with $\omega a/2\pi c_0 = 0.0211$ propagating along a convex-shaped interface in a finite system. The insert is the configuration of the structure. The blue line denotes the interface. The red arrows indicate the directions in which the cylinders rotate. The red stars mark the positions of point sources.